\documentclass[pra,superscriptaddress,amssymb]{revtex4}

\usepackage{amssymb}
\usepackage{graphicx}
\usepackage{dcolumn}
\usepackage{bm}
\usepackage{amsfonts,amssymb,amsmath}        % for math symbols.
\usepackage[a4paper,margin=1in]{geometry}
\usepackage{braket}
\usepackage{color}
\usepackage{sidecap}

\newcommand{\pp}{\boldsymbol{p}}
\newcommand{\PSI}{\boldsymbol{\psi}}
\newcommand{\noi}{\noindent}
\newcommand{\nn}{\nonumber}
\newcommand{\da}{\dagger}
\newcommand{\s}{\boldsymbol{\hat{s}}}
\newcommand{\SIG}{\boldsymbol{\sigma}}
\newcommand{\dBZ}{\partial BZ}
\newcommand{\PA}{\boldsymbol{\partial}}
\newcommand{\px}{\partial_{p_{x}}}
\newcommand{\py}{\partial_{p_{y}}}
\newcommand{\sv}{\boldsymbol{s}}
\newcommand{\SIGMA}{\boldsymbol{\Sigma}}
\newcommand{\bpm}{\begin{pmatrix}}
\newcommand{\epm}{\end{pmatrix}}
\newcommand{\jv}{\boldsymbol{j}}
\newcommand{\iv}{\boldsymbol{i}}
\newcommand{\x}{\boldsymbol{\hat{x}}}
\newcommand{\y}{\boldsymbol{\hat{y}}}
\newcommand{\rf}[1]{(\ref{#1})}

\begin{document}

\title{Detection of Chern numbers and entanglement in topological two-species systems through subsystem winding numbers}

\author{James de Lisle}
\affiliation{School of Physics and Astronomy, University of Leeds, Leeds, LS2 9JT, United Kingdom}
\author{Suvabrata De}
\affiliation{School of Physics and Astronomy, University of Leeds, Leeds, LS2 9JT, United Kingdom}
\author{Emilio Alba}
\affiliation{Instituto de F\'{\i}sica Fundamental, IFF-CSIC, Calle Serrano 113b, Madrid 28006, Spain} 
\author{Alex Bullivant}
\affiliation{School of Physics and Astronomy, University of Leeds, Leeds, LS2 9JT, United Kingdom}
\author{Juan J. Garcia-Ripoll}
\affiliation{Instituto de F\'{\i}sica Fundamental, IFF-CSIC, Calle Serrano 113b, Madrid 28006, Spain}
\author{Ville Lahtinen}
\affiliation{Institute for Theoretical Physics, University of Amsterdam, Science Park 904, NL-1090 GL, Amsterdam, The Netherlands}
\affiliation{Institute-Lorentz for Theoretical Physics, Leiden University, PO Box 9506, NL-2300 RA Leiden, The Netherlands}
\author{Jiannis K. Pachos}
\affiliation{School of Physics and Astronomy, University of Leeds, Leeds, LS2 9JT, United Kingdom}

\begin{abstract}

Topological invariants, such as the Chern number, characterise topological phases of matter. Here we provide a method to detect Chern numbers in systems with two distinct species of fermion, such as spins, orbitals or several atomic states. We analytically show that the Chern number can be decomposed as a sum of component specific winding numbers, which are themselves physically observable. We apply this method to two systems, the quantum spin Hall insulator and a staggered topological superconductor, and show that (spin) Chern numbers are accurately reproduced. The measurements required for constructing the component winding numbers also enable one to probe the entanglement spectrum with respect to component partitions. Our method is particularly suited to experiments with cold atoms in optical lattices where time-of-flight images can give direct access to the relevant observables.

\end{abstract}

\maketitle

\section{INTRODUCTION}

\noi With the discovery of the integer quantum Hall effect \cite{KDP80} physics saw a paradigm shift in the effort to understand the various phases of matter. This paradigm, which goes beyond the Landau symmetry breaking approach to the classification of different orders, has become known as topological order \cite{W95}. Research into topological order continues at great pace today, with the recent discovery of topological insulators and superconductors \cite{HK10,QZ11,B13}. This effort is spurred on by the prospect of using topological matter for quantum computation \cite{P12,NCS08}. 

The definitive theoretical signature of topological order is the existence of a topological invariant that characterises the ground state. While the exact form of the invariant depends on the symmetries of the system, one exists in general for all gapped systems. This has enabled the classification of all topologically ordered states of non-interacting fermions \cite{SRFL08,K09}. Of particular recent interest have been two-dimensional time-reversal symmetry broken systems characterised by Chern numbers \cite{TKNN82,K85}. These include (fractional) quantum Hall states, topological $p$-wave superconductors or (fractional) Chern insulators, which under suitable conditions are all predicted to support localised Majorana modes. Another topical class of systems are the experimentally accessible time-reversal symmetric topological insulators that are characterised by $\mathbb{Z}_2$ invariants \cite{Kane05}. 

While the topological invariants make the theoretical characterisation of topological quantum matter straightforward and unambiguous, it is not straightforward to determine them experimentally. Topological invariants are not in general related to readily measurable observables. Instead, the states tend to be characterised by secondary signatures such as the quantum Hall effect, the thermal Hall effect \cite{SF13,GD09,OSN08}, the zero-frequency conductivity \cite{H13}, the ground state degeneracy \cite{W95}, the topological entanglement entropy \cite{KP06,LW06}, the existence of edge states or the properties of the entanglement spectrum \cite{LH08}. These can function as definite smoking guns for topological order, but they are usually insufficient to provide the full characterisation that can be provided by the topological invariant.  

Due to their inherent clean nature, cold atoms in optical lattices promise an experimentally accessible route to prepare and detect different topological states of matter. Recently much progress has been made to detect topological characteristics in different physical observables available in such systems. These methods include time-of-flight measurements \cite{WST13,PALG13,AFMPG11,GAGOSJ11}, analysis of wave packet dynamics in optical lattices \cite{PC12}, interferometric measurements of the Berry phase \cite{AKBD13,AABAKDB12}, measurements of the Hall conductance \cite{BAR8,SZSXW08}, measurements of the centre of mass of atomic gases \cite{DG13} and direct measurements of the Skyrmion number \cite{BB13}. However, these schemes are often tailored for particular systems and require idealistic conditions that can be unrealistic in actual experiments.

In this work we generalise the method to detect Chern numbers in time-of-flight images \cite{WST13,PALG13,AFMPG11} -- a standard diagnostic technique in optical lattice experiments -- to two-species systems. To engineer the delicate interactions that give rise to topological order, such as spin-orbit coupling, non-Abelian gauge fields or $p$-wave pairing interactions, one often needs many species of atoms and/or internal states. This in turn tends to complicate the measurement schemes that usually rely on single species observables. Our main result is to show that such single species observables are sufficient to construct the full Chern number also in two-species systems. In addition, we show that such observables allow us to diagnose the degree of entanglement between different system components and access the entanglement spectrum with respect to  component bi-partitions \cite{LN13}.   
 
\subsection{Warm-up: Chern numbers and winding numbers in single component systems }
 
To introduce the basic concepts of our argument and to motivate for overcoming the problems which arise when the system grows in complexity, we start by briefly reviewing the detection of Chern numbers from time-of-flight images. Let us consider a translationally invariant non-interacting system of fermions in two spatial dimensions described by the Hamiltonian $H = \int_{\pp\in BZ}d^{2}p\,\PSI^{\da}_{\pp}h(\pp)\PSI_{\pp}$, where $BZ$ denotes the Brillouin zone. We assume that the system has two components, i.e. that the Hamiltonian is given in the basis $\PSI_{\pp} = (a_{\pp} , b_{\pp})^{T}$ for some fermionic operators $a_{\pp}$ and $b_{\pp}$. The ground state of the system is given by $\ket{\Psi}=\prod_{\pp}\ket{\psi_{\pp}}$ which is defined in terms of the fermionic operators $a^{\dagger}_{\pp}$ and $b^{\dagger}_{\pp}$ acting on the vacuum, $\ket{00}$. Furthermore, we require that the system has an energy gap. The kernel Hamiltonian $h(\pp)$ has two energy bands $E_{1}(\pp)\leq0$ and $E_{2}(\pp)\geq0$ such that if $\text{min}_{\pp}|E_{2}(\pp)|-\text{max}_{\pp}|E_{1}(\pp)|\neq0$ then the system has a gap. The Chern number characterising the ground state of such a system is formally defined as
\begin{equation}\label{eqn:chernproj}
\nu=-\frac{i}{2\pi}\int_{BZ}d^{2}p \,\,\text{tr}\left(P_{\pp}\left[\partial_{p_{x}}P_{\pp},\partial_{p_{y}}P_{\pp}\right]\right),
\end{equation}
where $P_{\pp}=\ket{\psi(\pp)}\bra{\psi(\pp)}$ is the projector onto the ground state of $h(\pp)$, $\ket{\psi(\pp)}$. Depending on how the projector is represented (see Appendix \ref{App:Chernreps} for detailed derivations), the Chern number can equivalently be expressed as the Berry phase accrued around the boundary $\dBZ$ of the two dimensional Brillouin zone 
\begin{equation}\label{berry}
\nu=-\frac{i}{2\pi}\oint_{\dBZ}\braket{\psi(\pp)|\PA|\psi(\pp)}\cdot d\pp,
\end{equation}
or as the winding number
\begin{equation}\label{equation:winding}
\nu=\frac{1}{4\pi}\int_{BZ}d^{2}p \,\, \s(\pp)\cdot\left(\px\s(\pp)\times\py\s(\pp)\right),
\end{equation}
that counts how many times the three-vector $\s(\pp)$ winds around the unit sphere as the Brillouin zone is spanned. This normalised vector parametrises the Hamiltonian through
\begin{equation}\label{equation:expansion}
h(\pp)=|\sv|\s(\pp)\cdot\SIG,
\end{equation}
where $\SIG =(\sigma^x,\sigma^y,\sigma^z)$ is the vector of Pauli matrices. It was shown in \cite{AFMPG11} for Chern insulators, where  $a_{\pp}$ and $b_{\pp}$ denote the two sublattices of the Haldane model on a honeycomb lattice, and in \cite{PALG13} for topological superconductors, where $b_{\pp}=a_{-\pp}^\dagger$, that the vector $\s(\pp)$ can be constructed from physical observables associated with time-of-flight measurements. As the representation (\ref{equation:winding}) of the Chern number is fully given by $\s(\pp)$, this means that the Chern number can be constructed from such measurements. 

To be more precise, by studying how the atom cloud expands as the trap is switched off, one obtains a set of time-of-flight images that amount to measuring density operators of the form $a_{\pp}^\dagger a_{\pp}$ and $b_{\pp}^\dagger b_{\pp}$. As these correspond to different species, one can assume that they can be measured independently by releasing only one species from the trap at a time. By direct evaluation one then finds that $\s(\pp)$ can be written either as the expectation value of the ground state of the Bloch Hamiltonian, $h_{\pp}$, with the Pauli matrices or as the expectation value of the ground state of the full Hamiltonian, $H$, with the Pauli operators in the second quantised representation
\begin{equation}\label{equation:s}
	\s(\pp)=\braket{\psi(\pp)|\SIG|\psi(\pp)}=\braket{\psi_{\pp}|\PSI^{\dagger}_{\pp}\SIG\PSI_{\pp}|\psi_{\pp}}.
\end{equation}
These two representations are equivalent. The $s_z$ component is thus given directly by the time-of-flight images, while the other components can be obtained by suitable Hamiltonian manipulations \cite{PALG13,AFMPG11}. 

This simple relation between physical observables and the Chern number breaks down when the number of degrees of freedom is increased, e.g. through the introduction of spin, staggering or more internal atomic states. The Bloch Hamiltonian, $h(\pp)$, is no longer a $2 \times 2$ matrix and can thus no longer be expanded in the Pauli basis $\{\sigma^x,\sigma^y,\sigma^z\}$. The Hamiltonian can still be expanded in a basis of some higher dimensional matrices and parametrised by a vector $\s(\pp)$, but this vector will now live in a space of dimension larger than three \cite{B13}. It follows that the formal definition of the Chern number (\ref{eqn:chernproj}) no longer reduces to a simple winding number (\ref{equation:winding}). While the components of the higher dimensional $\s(\pp)$ could still be in principle obtained via time-of-flight images following Hamiltonian manipulations, there is no longer a recipe for how the Chern number could be constructed from them. 

Our main result is to show that the Chern number can be decomposed into component dependent contributions, which can be evaluated as physically observable winding numbers \eqref{equation:winding}, as in the simple example described above. Our construction generalises the concept of spin Chern number to general pseudospin degrees of freedom. We demonstrate our method by applying it to spin Hall insulators \cite{Kane05} and a staggered topological superconductor \cite{PALG13}. In each case the phase diagrams are accurately reproduced given that the subsystem components are not highly entangled. This can be diagnosed using the same operators as the ones used to construct the subsystem winding numbers, which means that our method enables us to probe entanglement between different degrees of freedom.

%In this paper we consider systems with two spin degrees of freedom. We show that by implementing a Schmidt decomposition on the ground state it is possible to write the Chern number as a sum of Berry phases of each spin component. Moreover, we find that the expectation value of the ground state with the spin observables of each subspace produce vectors of the form \eqref{equation:s}. This allows us to associate, under certain conditions, a winding number with each spin component. The existence of a winding number for each subspace and the decomposition of the Berry phase allows us to write the Chern number as the sum of these winding numbers, both of which are observables.

\section{Decomposition of the Chern number into subsystem winding numbers}
\label{sec:decomposition}

In this section we give an analytic argument for decomposing the Chern number into a sum of winding numbers associated with the different components. A detailed derivation is first presented for particle number conserving topological insulators. The argument is then shown to apply equally to particle parity conserving topological superconductors.

\subsection{Decomposition for topological insulators}

Let us consider a system with four distinct types of fermion, whose annihilation operators we denote by $a_{1}$, $a_{2}$, $b_{1}$ and $b_{2}$, where the \textit{a priori} bipartition between $a$ and $b$ is physically motivated (such as different spin orientations, atomic internal levels or different sectors of some discrete symmetry). Assuming translational invariance with respect to these operators the Hamiltonian can always be given in  momentum space as
\begin{equation}\label{ham1}
H=\int_{\pp\in BZ}d^{2}p\,\Psi^{\da}_{\pp}h(\pp)\Psi_{\pp}\qquad \text{with}\qquad \Psi_{\pp}=\bpm a_{1,\pp}\\a_{2,\pp}\\b_{1,\pp}\\b_{2,\pp}\epm.
\end{equation}
A general state in the Hilbert space of the system can be written as
\begin{equation}\label{sta}
\ket{\Psi}=\prod_{\pp}\left(\sum_{n^{a}_{1,\pp},n^{a}_{2,\pp},n^{b}_{1,\pp},n^{b}_{2,\pp}=0,1}\alpha_{n^{a}_{1,\pp},n^{a}_{2,\pp},n^{b}_{1,\pp},n^{b}_{2,\pp}}\ket{n^{a}_{1,\pp},n^{a}_{2,\pp},n^{b}_{1,\pp},n^{b}_{2,\pp}}\right)\equiv\prod_{\pp}\ket{\psi_{\pp}},
\end{equation}
where we have expressed it in the occupation basis
\begin{equation}\label{eq:order}
\ket{n^{a}_{1,\pp},n^{a}_{2,\pp},n^{b}_{1,\pp},n^{b}_{2,\pp}}=(a_{1,\pp}^{\da})^{n^{a}_{1,\pp}}(a_{2,\pp}^{\da})^{n^{a}_{2,\pp}}(b_{1,\pp}^{\da})^{n^{b}_{1,\pp}}(b_{2,\pp}^{\da})^{n^{b}_{2,\pp}}\ket{0000}.
\end{equation}
Here $n^{a,b}_{i,\pp}=0,1$ are the fermionic occupation numbers and $\ket{0000}$ corresponds to the vacuum state of all fermionic modes. Eigenstates of Hamiltonian \eqref{ham1} will be of this form for some set of coefficients $\alpha_{n^{a}_{1,\pp},n^{a}_{2,\pp},n^{b}_{1,\pp}n^{b}_{2,\pp}}$ that satisfy $\sum_{n^{a}_{1,\pp},n^{a}_{2,\pp},n^{b}_{1,\pp}n^{b}_{2,\pp}=0,1}|\alpha_{n^{a}_{1,\pp},n^{a}_{2,\pp},n^{b}_{1,\pp}n^{b}_{2,\pp}}|^{2}=1$.

Let us now assume that the system conserves particle number, i.e. $[H,N]=0$, where $N=\sum_{\pp,\alpha=1,2}(a^{\da}_{\alpha,\pp}a_{\alpha,\pp}+b^{\da}_{\alpha,\pp}b_{\alpha,\pp})$. At half filling the ground state $\ket{\psi_{\pp}}$ satisfies the condition $\sum_{\alpha=1,2}(n^{a}_{\alpha,\pp}+n^{b}_{\alpha,\pp})=2$. This means that a complete local basis for each momentum component of the ground state is given by
\begin{equation}
\left\{\ket{1100},\ket{1010},\ket{1001},\ket{0110},\ket{0101},\ket{0011}\right\}.
\end{equation}
We can divide the state into two orthogonal subspaces
\begin{eqnarray}\label{eo}
\ket{\psi_{\pp}}&=&A\ket{\psi(n^{a}_{1,\pp}+n^{a}_{2,\pp}=\text{even};n^{b}_{1,\pp}+n^{b}_{2,\pp}=\text{even})}+B\ket{\psi(n^{a}_{1,\pp}+n^{a}_{2,\pp}=\text{odd};n^{b}_{1,\pp}+n^{b}_{2,\pp}=\text{odd})}\nn\\
&\equiv& A\ket{\psi(e;e)}+B\ket{\psi(o;o)},
\end{eqnarray}
where the populations $n^{a}_{1,\pp}+n^{a}_{2,\pp}$ and $n^{b}_{1,\pp}+n^{b}_{2,\pp}$ are either both even or both odd for some $A$ and $B$, with $|A|^{2}+|B|^{2}=1$. This partitioning of the state in this particular way facilitates our derivation. 

We now perform a Schmidt decomposition on each part of the state, the even and the odd. We can write
\begin{eqnarray}\label{equation:schmidt}
\ket{\psi(e;e)}&=&\cos\theta_{e}\ket{a_{e}}\otimes\ket{b_{e}}+\sin\theta_{e}\ket{\tilde{a}_{e}}\otimes\ket{\tilde{b}_{e}},\nn\\
\ket{\psi(o;o)}&=&\cos\theta_{o}\ket{a_{o}}\otimes\ket{b_{o}}+\sin\theta_{o}\ket{\tilde{a}_{o}}\otimes\ket{\tilde{b}_{o}},
\end{eqnarray} 
where $\theta_{e},\theta_{o}\in[0,\pi/2)$ such that all the Schmidt coefficients are non-negative. It is understood that the $\ket{a_{e/o}}\ket{b_{e/o}}$ states written in the population basis have the fermionic operators ordered as in \eqref{eq:order}. The vectors $\ket{\tilde{a}_{e/o}}\ket{\tilde{b}_{e/o}}$ obey the orthogonality conditions $\braket{a_{e/o}|\tilde{a}_{e/o}}=0$ and $\braket{b_{e/o}|\tilde{b}_{e/o}}=0$. More explicitly we have
\begin{alignat}{3} \label{subsystems}
\ket{a_{o}}&=\left(\alpha_{01}a^{\da}_{2,\pp}+\alpha_{10}a^{\da}_{1,\pp}\right)\ket{00}, \quad &\ket{\tilde{a}_{o}}&=\left(\alpha^{*}_{10}a^{\da}_{2,\pp}-\alpha^{*}_{01}a^{\da}_{1,\pp}\right)\ket{00},\nn\\
\ket{b_{o}}&=\left(\beta_{01}b^{\da}_{2,\pp}+\beta_{10}b^{\da}_{1,\pp}\right)\ket{00}, \quad &\ket{\tilde{b}_{o}}&=\left(\beta^{*}_{10}b^{\da}_{2,\pp}-\beta^{*}_{01}b^{\da}_{1,\pp}\right)\ket{00}
\end{alignat}
and
\begin{alignat}{3}
\ket{a_{e}}&=e^{i\phi_{a}}\ket{00}, \quad &\ket{\tilde{a}_{e}}&=e^{i\tilde{\phi}_{a}}a_{1,\pp}^{\da}a_{2,\pp}^{\da}\ket{00},\nn\\
\ket{b_{e}}&=e^{i\phi_{b}}b_{1,\pp}^{\da}b_{2,\pp}^{\da}\ket{00}, \quad &\ket{\tilde{b}_{e}}&=e^{i\tilde{\phi}_{b}}\ket{00},
\end{alignat}
where $|\alpha_{01}|^{2}+|\alpha_{10}|^{2}=|\beta_{01}|^{2}+|\beta_{10}|^{2}=1$. The phases $\phi_{a/b}$ and $\tilde{\phi}_{a/b}$ are in general non-zero. However, after multiplying $\ket{\psi_{\pp}}$ by a global phase of $e^{-i(\phi_{a}+\phi_{b})}$, we can transfer them to the odd-odd subspace through a $U(1)$ gauge transformation $a_{1,\pp}^{\da}\rightarrow e^{-i(\tilde{\phi}_{a}+\tilde{\phi}_{b}-\phi_{a}-\phi_{b})}a_{1,\pp}^{\da}$. Then the only momentum dependence in the Schmidt decomposed even-even subspace is in the real and positive Schmidt coefficients, $\cos\theta_{e}$ and $\sin\theta_{e}$.

We now show that we can write the Chern number of the system as a sum of the Berry phases accrued by each subsystem. To this end, we evaluate the Chern number of the ground state of the system, \eqref{eo}, as the Berry phase \eqref{berry}. Without loss of generality we can take $A$ and $B$ to be real and non-negative. This is achieved by absorbing possible complex phases into the states $\ket{a_{e/o}}\ket{b_{e/o}}$ and $\ket{\tilde{a}_{e/o}}\ket{\tilde{b}_{e/o}}$. Employing relation \eqref{eo} we can write the Chern number as
\begin{equation}\label{berry1}
\nu=-\frac{i}{2\pi}\oint_{\dBZ}A^{2}\braket{\psi(e;e)|\PA|\psi(e;e)}\cdot d\pp -\frac{i}{2\pi}\oint_{\dBZ}B^{2}\braket{\psi(o;o)|\PA|\psi(o;o)}\cdot d\pp,
\end{equation}
where terms of the form $A\PA A$ do not contribute as, due to the reality condition on $A$ and $B$, $A\PA A+B\PA B=\PA (A^{2}+B^{2})/2=0$. Direct evaluation of the integrand in the even-even case finds it to be zero due to $\cos\theta_{e}\PA\cos\theta_{e}+\sin\theta_{e}\PA\sin\theta_{e}=\PA(\cos^{2}\theta_{e}+\sin^{2}\theta_{e})/2=0$. Thus the first term on the right hand side of \eqref{berry1} does not contribute to the Berry phase. Noting that $\braket{i_{o}|\PA|i_{o}}=-\braket{\tilde{i}_{o}|\PA|\tilde{i}_{o}}$ in (\ref{subsystems}) and using the positivity and normalisation of the Schmidt coefficients, a direct evaluation gives
\begin{equation}\label{equation:suberry}
\nu=-\frac{i}{2\pi}\sum_{i=a,b}\oint_{\dBZ}S\braket{i_{o}|\PA|i_{o}}\cdot d\pp,\qquad S=|B|^{2}T,
\end{equation}
where $T=\cos^{2}\theta_{o}-\sin^{2}\theta_{o}$ is a measure of entanglement between the subsystems.

Thus we have succeeded in decomposing the Chern number into a sum of exclusive contributions from the $a$ or $b$ subsystems. Due to appearance of the function $S$, these contributions are not Berry phases which one could equivalently evaluate as winding numbers of vectors $\s_{a}(\pp)$ and $\s_{b}(\pp)$. However, the decomposition holds for any $S\neq0$. For $S=1$ the two subsystems are unentangled and the Chern number can be directly written as a sum of Berry phase contributions. But the decomposition \emph{also} holds for $S\neq1$ and fails only when $S\rightarrow0$ where the system is maximally entangled. This can be seen in the examples of Section \ref{sec:numerics} where our method diverges from the theoretical predictions only when the system approaches maximal entanglement.

\subsection{Subsystem winding numbers as physical observables}

In direct analogy with the single component case presented in the introduction, we define the observables for the $a$ and $b$ subsystems as
\begin{equation} \label{sub_obs}
\begin{array}{rclrclrcl}
\Sigma_{a}^{x} & = & a_{1,\pp}^{\da}a_{2,\pp}+a_{2,\pp}^{\da}a_{1,\pp}, & \Sigma_{a}^{y} & = & -ia_{1,\pp}^{\da}a_{2,\pp}+ia_{2,\pp}^{\da}a_{1,\pp}, & \Sigma_{a}^{z} & = & a_{1,\pp}^{\da}a_{1,\pp}-a_{2,\pp}^{\da}a_{2,\pp} \\
\Sigma_{b}^{x} & =& b_{1,\pp}^{\da}b_{2,\pp}+b_{2,\pp}^{\da}b_{1,\pp}, & \Sigma_{b}^{y} & = & -ib_{1,\pp}^{\da}b_{2,\pp}+ib_{2,\pp}^{\da}b_{1,\pp}, & \Sigma_{b}^{z} & =& b_{1,\pp}^{\da}b_{1,\pp}-b_{2,\pp}^{\da}b_{2,\pp}.
\end{array}
\end{equation}
Since these operators conserve the number of particles there are no cross-terms between the even and odd subspaces of \eqref{eo}, and thus the expectation values of these operators $\SIGMA_{i,\pp}=\bpm\Sigma_{i}^{x},\Sigma_{i}^{y},\Sigma_{i}^{z}\epm$ read
\begin{equation}\label{equation:expa}
\braket{\psi_{\pp}|\SIGMA_{i,p}|\psi_{\pp}}=|A|^{2}\braket{\psi(e;e)|\SIGMA_{i,p}|\psi(e;e)}+|B|^{2}\braket{\psi(o;o)|\SIGMA_{i,p}|\psi(o;o)}.
\end{equation}
By direct evaluation we find that the contribution from the even subspace vanishes. On the other hand, the odd subspace component gives
\begin{equation}\label{ev2}
\braket{\psi(o;o)|\SIGMA_{i,\pp}|\psi(o;o)}=\cos^{2}\theta_{o}\braket{i_{o}|\SIGMA_{i,\pp}|i_{o}}+\sin^{2}\theta_{o}\braket{\tilde{i}_{o}|\SIGMA_{i,\pp}|\tilde{i}_{o}}=T\braket{i_{o}|\SIGMA_{i,\pp}|i_{o}},
\end{equation}
where we have used the tracelessness of the $\SIGMA_{i,\pp}$ operators that implies $\braket{\tilde{i}_{o}|\SIGMA_{i,\pp}|\tilde{i}_{o}}=-\braket{i_{o}|\SIGMA_{i,\pp}|i_{o}}$. Altogether we then obtain
\begin{eqnarray}\label{equation:exp1}
\braket{\psi_{\pp}|\SIGMA_{a,\pp}|\psi_{\pp}}& = & S\braket{\psi_{a}(\pp)| \SIG |\psi_{a}(\pp)}, \\
\braket{\psi_{\pp}|\SIGMA_{b,\pp}|\psi_{\pp}}& = & S\braket{\psi_{b}(\pp)| \SIG |\psi_{b}(\pp)} \nonumber
\end{eqnarray}
where $\ket{\psi_{a}(\pp)}=\bpm\alpha_{01},&\alpha_{10}\epm^{T}$ and $\ket{\psi_{b}(\pp)}=\bpm\beta_{01},&\beta_{10}\epm^{T}$. In other words, we obtain two three-vectors $\s_{a}(\pp)$ and $\s_{b}(\pp)$ whose normalized components are given by
\begin{equation}\label{equation:sd}
\s_{i}(\pp)=\frac{\sv_{i}(\pp)}{|\sv_{i}(\pp)|}=\braket{\psi_{i}(\pp)|\SIG|\psi_{i}(\pp)}, \qquad |\sv_{i}(\pp)|=|B|^{2}|T|.
\end{equation}
The fact that the norm of these vectors is equal to $|S|$ implies that the degree of entanglement between the subsystems can be probed by using the operators \rf{sub_obs}.

Exactly like the Berry phase \eqref{berry} and the winding number \eqref{equation:winding} were equivalent representations of the Chern number for two component systems, we can define a subsystem Chern number $\nu_i$ for the state $\ket{i_o}$ and represent it equivalently either as the subsystem Berry phase
\begin{equation}
	\nu_i=\frac{i}{2\pi} \oint_{\dBZ}\braket{i_{o}|\PA|i_{o}}\cdot d\pp,
	\label{eqn:sberry}
\end{equation}
or as the subsystem winding numbers 
\begin{equation} \label{subwinding}
\nu_i=\frac{1}{4\pi}\int_{BZ}d^{2}p \,\, \s_i(\pp)\cdot\left(\px\s_i(\pp)\times\py\s_i(\pp)\right).
\end{equation}
Formally this comes about by viewing $\ket{\psi_{i}(\pp)}$ as the ground state of a fictitious Hamiltonian $H_i=\s_{i}(\pp)\cdot \SIG$ with eigenvalues $E=\pm 1$. As we have shown above, the components of the vectors $\s_{i}(\pp)$ can be obtained from the observables \rf{sub_obs} and hence the subsystem Chern number is an observable.

We have shown that each subsystem Berry phase \eqref{eqn:sberry} is proportional to a winding number \eqref{subwinding} that is physically observable. When $S=1$, by \eqref{equation:suberry}, the Chern number becomes additive in the subsystem winding numbers. However, the method also works for $S\neq1$ and examples in Section \ref{sec:numerics} show that the Chern number is returned as long as the system is not maximally entangled, or $S\nrightarrow0$. We will now show that with small modifications the same method applies directly also to topological superconductors that conserve only the particle parity.

%From the existence of the vectors defined in \eqref{equation:sd} we can associate a winding number with each subspace, $i=a,b$. Given a winding number for some state of a system we know that formally there exists a corresponding Berry phase. Subsequently, we show numerically that the sum of these winding numbers exactly reproduces the Chern number for our examples.  Indeed, if we take $S\rightarrow1$ the Chern number decomposition \eqref{equation:suberry} becomes a sum of the winding numbers of each fermion species. Theoretically, this is possible when the continuous transition $S\rightarrow1$ can happen without going through a phase transition. This would not be possible, e.g. when there is a parametric regime for which $S=0$, i.e. when the Schmidt decomposition \eqref{equation:schmidt} corresponds to maximally entangled states. This behaviour becomes apparent in the examples discussed below.

\subsection{Decomposition for topological superconductors}
\label{sec:spinfulsuperconductors}

The generalisation to topological superconductors is straightforward. We take the Hamiltonian to be of the same form as \eqref{ham1} with the basis given now by $\Psi^{\da}_{\pp}=\bpm a^{\da}_{\pp}, a_{-\pp}, b^{\da}_{\pp},b_{-\pp}\epm$. The general state can be written as \eqref{sta} with the Fock space ordered as
\begin{equation}
\ket{n^{a}_{\pp},n^{a}_{-\pp},n^{b}_{\pp},n^{b}_{-\pp}}=(a_{\pp}^{\da})^{n^{a}_{\pp}}(a_{-\pp}^{\da})^{n^{a}_{-\pp}}(b_{\pp}^{\da})^{n^{b}_{\pp}}(b_{-\pp}^{\da})^{n^{b}_{-\pp}}\ket{0000}.
\end{equation}
A superconducting system conserves only the total parity, i.e. $\left[H,P\right]=0$ with $P=\exp\left(i\pi\sum_{\pp}\left(a_{\pp}^{\da}a_{\pp}+b_{\pp}^{\da}b_{\pp}\right)\right)=P_{a}P_{b}$, while component parities $P_{a}$ and $P_{b}$ are not independently conserved. Without loss of generality we assume that the ground state resides in the even total parity sector. This means parities in the subsystems $a$ and $b$ are correlated such that $P_{a}=P_{b}$, which in turn means that the ground state complies with the condition of zero overall momentum. In this parity sector the ground state is thus given in the basis spanned by the states
\begin{equation}
\left\{\ket{0000},\ket{0011},\ket{1100},\ket{1111},\ket{0110},\ket{1001}\right\}.
\end{equation}

As with the topological insulators, we split the Hilbert space into even and odd occupation subspaces and write 
\begin{equation}\label{equation:ss} 
\ket{\psi_{\pp}}=A\ket{\psi(e;e)}+B\ket{\psi(o;o)}.
\end{equation}
Performing a Schmidt decomposition between the subsystems $a$ and $b$ in this parity sector, we obtain a general expression which has the same form as \eqref{equation:schmidt}, but with the Schmidt bases now being given by
\begin{alignat}{3}
\ket{a_{e}}&=\left(\alpha_{00}+\alpha_{11}a^{\da}_{\pp}a^{\da}_{-\pp}\right)\ket{00}, \quad &\ket{\tilde{a}_{e}}&=\left(\alpha^{*}_{11}-\alpha^{*}_{00}a^{\da}_{\pp}a^{\da}_{-\pp}\right)\ket{00},\nn\\
\ket{b_{e}}&=\left(\beta_{00}+\beta_{11}b^{\da}_{\pp}b^{\da}_{-\pp}\right)\ket{00}, \quad &\ket{\tilde{b}_{e}}&=\left(\beta^{*}_{11}-\beta^{*}_{00}b^{\da}_{\pp}b^{\da}_{-\pp}\right)\ket{00},
\end{alignat}
and
\begin{alignat}{3}
\ket{a_{o}}&=e^{i\phi_{a}}a^{\da}_{-\pp}\ket{00}, \quad &\ket{\tilde{a}_{o}}&=e^{i\tilde{\phi}_{a}}a^{\da}_{\pp}\ket{00},\nn\\
\ket{b_{o}}&=e^{i\phi_{b}}b^{\da}_{\pp}\ket{00}, \quad &\ket{\tilde{b}_{o}}&=e^{i\tilde{\phi}_{b}}b^{\da}_{-\pp}\ket{00}.
\end{alignat}
As in the case of topological insulators where all coefficients except those in the odd subspace could be made real through $U(1)$ gauge transformations, we can now take only the coefficients in the even subspace to be complex and all other coefficients to be real. The decomposition of the Berry phase proceeds in similar steps to the insulating case. The only difference is that it is now the odd subspace contribution that vanishes in \eqref{berry1}, with the Chern number being now given by
\begin{equation}\label{equation:berryds}
\nu=-\frac{i}{2\pi}\sum_{i=a,b}\oint_{\dBZ}S\braket{i_{e}|\PA|i_{e}}\cdot d\pp,\qquad S=|A|^{2}T.
\end{equation}

The relevant observables $\SIGMA_{i,\pp}=\bpm\Sigma_{i}^{x},\Sigma_{i}^{y},\Sigma_{i}^{z}\epm$ to evaluate the subsystem winding numbers are now defined by
\begin{equation}\label{eq:supobs}
\begin{array}{rclrclrcl}
\Sigma_{a}^{x} & =& a_{\pp}^{\da}a^{\da}_{-\pp}+a_{-\pp}a_{\pp}, & \Sigma_{a}^{y} & = & -ia_{\pp}^{\da}a^{\da}_{-\pp}+ia_{-\pp}a_{\pp}, &  \Sigma_{a}^{z} & = & a_{\pp}^{\da}a_{\pp}-a_{-\pp}^{\da}a_{-\pp}, \\
\Sigma_{b}^{x} & = & b_{\pp}^{\da}b^{\da}_{-\pp}+b_{-\pp}b_{\pp}, & \Sigma_{b}^{y} & = & -ib_{\pp}^{\da}b^{\da}_{-\pp}+ib_{-\pp}b_{\pp}, & \Sigma_{b}^{z} & = & b_{\pp}^{\da}b_{\pp}-b_{-\pp}^{\da}b_{-\pp}.
\end{array}
\end{equation}
Computing their expectation values we find that now only the even subspace contributes. The precise expressions are given by
\begin{eqnarray}\label{equation:expb}
\braket{\psi(e;e)|\SIGMA_{a,\pp}|\psi(e;e)} & =& S \braket{\psi_{a}(\pp)| \SIG |\psi_{a}(\pp)}, \\
\label{equation:expb1}\braket{\psi(e;e)|\SIGMA_{b,\pp}|\psi(e;e)} & =& S \braket{\psi_{a}(\pp)| \SIG |\psi_{a}(\pp)},
\end{eqnarray}
where now $\ket{\psi_{a}(\pp)}=\bpm\alpha_{00},&\alpha_{11}\epm^{T}$ and $\ket{\psi_{b}(\pp)}=\bpm\beta_{00},&\beta_{11}\epm^{T}$. Thus the expectation values can again be used to define two three-vectors  $\s_{a}(\pp)$ and $\s_{b}(\pp)$, that can be used to evaluate the subsystem winding numbers \rf{subwinding}. Under the same assumption of non-maximal entanglement between the subsystems, the Chern number will be shown to be additive in these winding numbers.

\section{Detection of subsystem entanglement spectrum}\label{secent}

We found above that the observable $\vert {\bf{s}}_i ({\bf{p}}) \vert \propto |\cos^{2}\theta_{{\bf{p}}}-\sin^{2}\theta_{{\bf{p}}}|$, i.e. that it provides a measure of entanglement between the subsystems. For $|{\bf{s}}_i ({\bf{p}})| \to 0$ the subsystems become maximally entangled, while for $\cos \theta_{{\bf{p}}} \to 1$ or $\sin \theta_{{\bf{p}}} \to 1$ the ground state becomes a product state. In fact, one can go further and use these same observables to construct the entanglement spectrum corresponding to component partitioning of the system that preserves  translational symmetry \cite{LN13}.

As was first pointed out by Li and Haldane \cite{LH08}, ground states described by reduced density matrices
\begin{equation}
  \rho_a = {\rm tr}_b \rho \propto e^{-\mathcal{H}_E^a}
\end{equation}
contain additional information if one considers the full spectrum of the entanglement Hamiltonian $\mathcal{H}_E^a$.  In the case of free or paired fermion problems, these are known to inherit the structure of the physical Hamiltonians in the sense that both can be formally written in the same basis \cite{P03}. Thus, the  entanglement Hamiltonians can be readily diagonalized with their eigenvalues $\epsilon_i$ constituting the (single particle) entanglement spectrum. We now show that these can be obtained directly from the observables $|{\bf{s}}_i (\bf{p})|$ for free fermion problems.

It has been shown by Peschel \cite{P03} that the entanglement energies $\epsilon_i$ of insulators can be obtained from ground state correlation functions $C_{ij}^{a}=\langle a_i^\dagger a_j \rangle$. To be precise, the eigenvalues $\lambda_i$ of this correlation matrix $\hat{C}^a$ are related to them through
\begin{equation} \label{entC}
	\lambda_i = (e^{\epsilon_i}+1)^{-1}.
\end{equation}
For every momentum component of the ground state $\hat{C}^a$ is a $2 \times 2$ matrix. It is easy to see that e.g. $s^z_a ({\bf p})=C^a_{11}-C^a_{22}$, and similarly for the $x$- and $y$-components. By direct calculation one then obtains the eigenvalues $\lambda_\pm = (N_a \pm |{\bf{s}}_a ({\bf{p}})|)/2$, where we have defined the occupation in the subsystem $a$ as $N_a = \langle a_1^\dagger a_1 + a_2^\dagger a_2 \rangle$ . Substituting this into \eqref{entC} yields the entanglement spectrum at each momenta
\begin{equation}
	\epsilon_\pm  ({\bf p}) = \ln \left( \frac{N_b \mp |{\bf{s}}_a (\bf{p})|}{N_a \pm |{\bf{s}}_a (\bf{p})|} \right),
\end{equation}
where $N_b = 2-N_a$ as we consider systems at half-filling. Hence, our observables give direct access also to the component entanglement spectrum studied in \cite{LN13}. The entanglement gap closes if $\epsilon_+ = \epsilon_-$ for some momentum mode $\bf{p}$. It is straighforward to verify that this is satisfied only when $ |{\bf{s}}_a ({\bf{p}})|=0$, i.e. when the subsystems are maximally entangled and our detection scheme becomes unreliable.

While a similar analytic derivation between the entanglement spectrum and the observables is more involved for paired fermion systems due to \eqref{entC} being replaced by a more complicated relation \cite{P03}, one can qualitatively understand that a similar relation must also hold for such systems. The ${\bf{p}}$-th component of $\rho_a$ is

\begin{equation}
\rho_a ({\bf{p}}) = |A|^2 \left( \cos^2 \theta_e \ket{a_e} \bra{a_e} + \sin^2 \theta_e \ket{\tilde{a}_e} \bra{\tilde{a}_e} \right) + |B|^2 \left( \cos^2 \theta_o \ket{a_o} \bra{a_o} + \sin^2 \theta_o \ket{\tilde{a}_o} \bra{\tilde{a}_o} \right),
\end{equation}
with $\rho_a$ being a product of these components over ${\bf{p}}$. The state with greatest weight in $\rho_a$ is the groundstate of the entanglement Hamiltonian, and since this Hamiltonian is of superconducting form, the groundstate must reside in the even parity sector. This means that the largest eigenvalue of $\rho_a$ is

\begin{equation}
\prod_{\bf{p}} |A|^2 {\rm max}(\cos^2 \theta_e , \sin^2 \theta_e).
\end{equation}
$\vert {\bf{s}}_a ({\bf{p}}) \vert \propto \vert \cos^2 \theta_e - \sin^2 \theta_e \vert$ vanishes when $\cos^2 \theta_e = \sin^2 \theta_e$, and when this happens, the largest eigenvalue of $\rho_a$ becomes degenerate. In other words, as in the case for insulating systems, when the subsystems are maximally entangled, the entanglement gap closes. Indeed, we will numerically show below that the observable $|{\bf{s}}_a (\bf{p})|$ and the entanglement gap are in exact agreement also for superconducting systems.

\section{CASE STUDIES}
\label{sec:numerics}

We now turn to demonstrate the validity of our analytic arguments for detecting Chern numbers through subsystem winding numbers by apply our scheme to two microscopically distinct examples: the quantum spin Hall insulator \cite{Kane05} and a staggered topological superconductor \cite{PALG13}. We show that in both cases the phase diagrams are accurately reproduced, with any discrepancies being attributable to high entanglement between the subsystems.

\subsection{Example I: The Quantum Spin Hall Insulator}

As the first example, we consider the quantum spin-Hall insulator defined on a honeycomb lattice \cite{Kane05}. The Hamiltonian of this model is given by
\begin{align}\label{eq:KM}
H =\, &t\sum_{\langle \iv \jv \rangle}a^\dagger_{\iv} b_{\jv}+\lambda_v\sum_{\iv} \left( a^\dagger_{\iv} a_{\iv}- b^\dagger_{\iv} b_{\iv} \right)\nn\\
&+ i \lambda_{SO}\sum_{\langle\langle \iv \jv \rangle\rangle} \xi_{SO}\left( a^\dagger_{\iv} \sigma_z a_{\jv}+ b^\dagger_{\iv} \sigma_z b_{\jv} \right) + i \lambda_{R}\sum_{\langle \iv \jv \rangle}a^\dagger_{\iv} \left(\boldsymbol{\sigma} \times \boldsymbol{\hat{d}}_{\iv}\right)_z b_{\jv}.
\end{align}
where the spinors $a^\dagger_{\iv} =(a^\dagger_{\iv,\uparrow},a^\dagger_{\iv,\downarrow})$ and $b^\dagger_{\iv} =(b^\dagger_{\iv,\uparrow},b^\dagger_{\iv,\downarrow})$ denote the two sublattice degrees of freedom of the honeycomb lattice. The first two terms of magnitudes $t$ and $\lambda_v$ describe spin-independent nearest-neighbour tunnelling and a sublattice energy imbalance, respectively. The other two terms proportional to $\lambda_R$ and $\lambda_{SO}$ are nearest and next-nearest neighbour spin-orbit couplings, respectively. In the notation used  $\xi_{SO}= \text{sign} (\boldsymbol{\hat{d}}_1 \times \boldsymbol{\hat{d}}_2)$ with $\boldsymbol{\hat{d}}_1$ and $\boldsymbol{\hat{d}}_2$ being vectors that connect the next-to-nearest neighbour sites. 

\begin{figure}[h!]
	\centering	
	\includegraphics[width=0.82\textwidth]{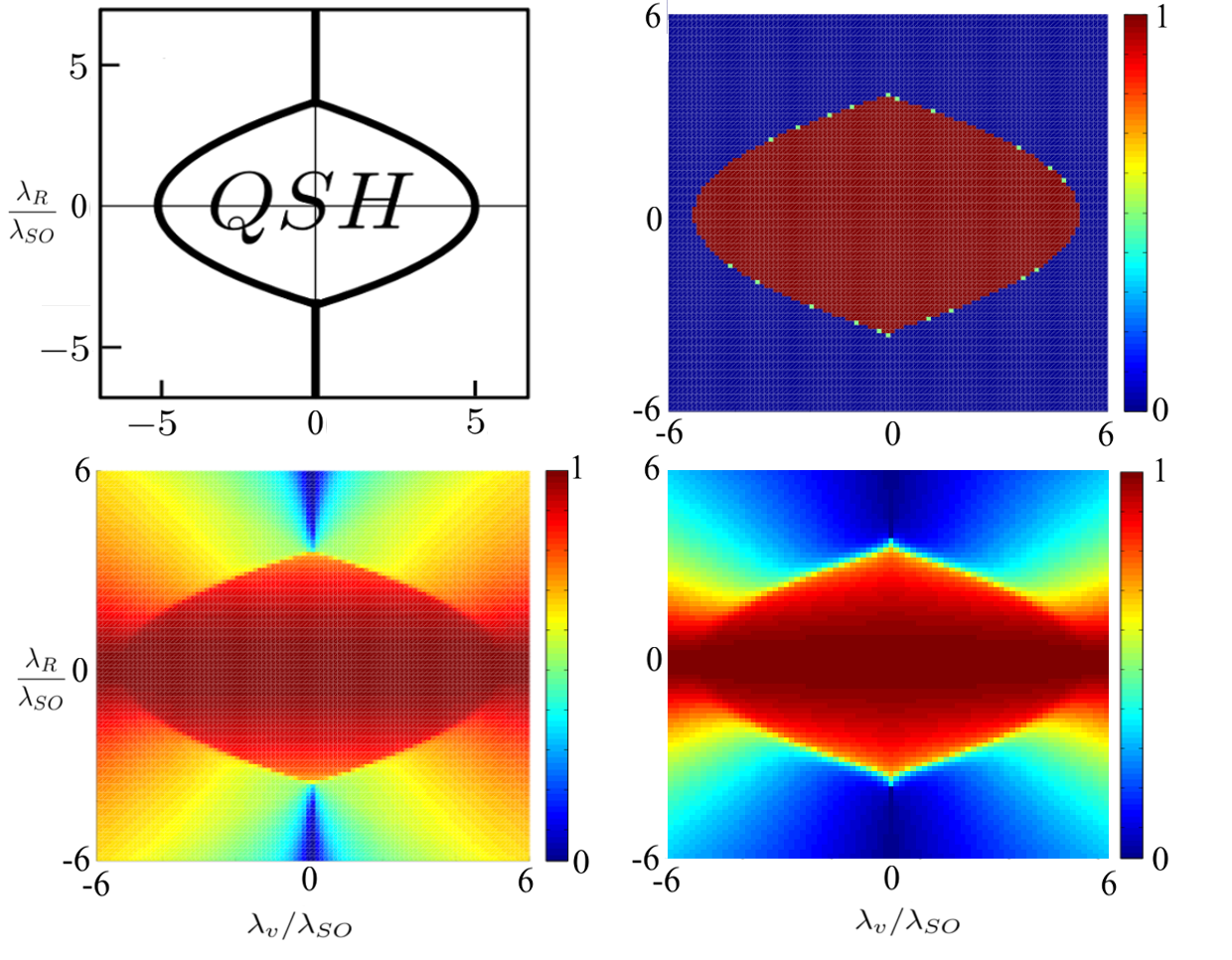}
	\caption{\emph{Top Left}: Theoretical phase diagram in the parameter-space $\lambda_R/\lambda_{SO}$, $\lambda_V/\lambda_{SO}$ of Hamiltonian \eqref{eq:KM}. The trivial phase corresponds to $\nu_S=0$, while the QSH phase corresponds to $\nu_S=1$. \emph{Top Right}: Numerical computation of the phase diagram as the winding spin Chern number $(\tilde{\nu}_{\uparrow}-\tilde{\nu}_{\downarrow})/2$. \emph{Bottom Left}: The minimum of the spin component entanglement measure across the Brillouin zone, $\text{min}_{\pp}|\sv_{i}(\pp)|$. When $\text{min}_{\pp}|\sv_{i}(\pp)|$ is small the entanglement is large. When We find the spin components becoming maximally entangled only around the transitions between the two trivial insulators, while between trivial and spin Hall phases we find a discontinuity. \emph{Bottom Right}: The gap of the entanglement spectrum corresponding to the spin up subsystem. This gap is defined to be ${\rm min}_{\pp} \lambda_+ - {\rm max}_{\pp} \lambda_-$ and is seen to close when $\sv_{i}(\pp)=0$, in agreement with the theoretical arguments presented in section \ref{secent}.}
	\label{fig:KM}
\end{figure}

By Fourier transforming the Hamiltonian \eqref{eq:KM} it takes the Bloch form \eqref{ham1} in the basis $\Psi_{\pp}=(a_{\uparrow,\pp},a_{\downarrow,\pp},b_{\uparrow,\pp},b_{\downarrow,\pp})^T$. By diagonalising this Hamiltonian Kane and Mele showed that it supports a trivial insulator and a quantum spin Hall phase, which are distinguished by a $\mathbb{Z}_2$ valued topological invariant \cite{Kane05}. While in a time-reversal symmetric system the Chern number is zero in all phases, the $\mathbb{Z}_2$ invariant was shown to be related to the so called spin Chern numbers that are quantised for each spin component \cite{SWSH06}. More precisely, the $\mathbb{Z}_2$ invariant was defined as the difference of the spin Chern numbers, $\nu_{S}=(\tilde{\nu}_{\uparrow}-\tilde{\nu}_{\downarrow})/2$, that only takes non-zero value in the quantum spin Hall phase. The phase diagram as a function of the microscopic parameters is shown in Fig. \ref{fig:KM}. 

The spin Chern number has a natural counterpart in our constrution if we identify the spin up and spin down components as the two subsystems with respect to which the ground state is Schmidt decomposed. The corresponding operators for evaluating the subsystem winding numbers are
\begin{equation} \label{sub_obs1}
\begin{array}{rclrclrcl}
\Sigma_{\uparrow}^{x} & = & a_{\uparrow,\pp}^{\da}b_{\uparrow,\pp}+b_{\uparrow,\pp}^{\da}a_{\uparrow,\pp}, & \Sigma_{\uparrow}^{y} & = & -ia_{\uparrow,\pp}^{\da}b_{\uparrow,\pp}+ib_{\uparrow,\pp}^{\da}a_{\uparrow,\pp}, & \Sigma_{\uparrow}^{z} & = & a_{\uparrow,\pp}^{\da}a_{\uparrow,\pp}-b_{\uparrow,\pp}^{\da}b_{\uparrow,\pp} 
\end{array}
\end{equation}
and similarly for the $\downarrow$-spin component. We construct the vectors $\s_{\uparrow}(\pp)$ and $\s_{\downarrow}(\pp)$ from these observables and by inserting them into \rf{subwinding}, calculate the corresponding subsystem winding numbers $\nu_\uparrow$ and $\nu_\downarrow$. We find that these observables give then precisely the spin Chern numbers $\tilde{\nu}_\uparrow$ and $\tilde{\nu}_\downarrow$, with Figure \ref{fig:KM} showing that the phase diagram is precisely reproduced. The figure also shows that the subsystem entanglement measure $|S|$ remains large within the QSH phase, which confirms that the spin components are minimally entangled in this phase. We take this as confirming the reliability of our method for non-maximally entangled states.

\subsection{Example II: Topological Superconductor with staggered sublattices}

The second example we consider is a recently introduced topological superconductor with a staggered chemical potential \cite{PALG13} that enables the system to support topological phases characterised by Chern numbers $\nu=0,\pm1$ and $\pm2$. The model is defined on a square lattice by the Hamiltonian
\begin{align}
H=\sum_{\jv}\Bigl[&(\mu-\delta)a_{\jv}^{\da}a_{\jv}+(\mu+\delta)b_{\jv}^{\da}b_{\jv}+t\Bigl(ia_{\jv}^{\da}b_{\jv}-ib_{\jv}^{\da}a_{\jv+\x}+a_{\jv}^{\da}a_{\jv+\y}+b_{\jv}^{\da}b_{\jv+\y}\Bigr)\nn\\
&+\Delta\Bigl(a_{\jv}^{\da}b_{\jv}^{\da}+b_{\jv}^{\da}a_{\jv+\x}^{\da}+a_{\jv}^{\da}a_{\jv+\y}^{\da}+b_{\jv}^{\da}b_{\jv+\y}^{\da}\Bigr)+h.c.\Bigr],
\end{align}
where $a^{\da}_{\jv}$ and $b^{\da}_{\jv}$ denote the two sublattice degrees of freedom that are distinguished by the staggered offset $\delta$ in the chemical potential $\mu$ and $\delta$. The other two coefficients $t$ and $\Delta$ correspond to the nearest neighbour hopping and pairing, respectively. 

\begin{figure}[h!]
	\center	
	\includegraphics[width=0.8\textwidth]{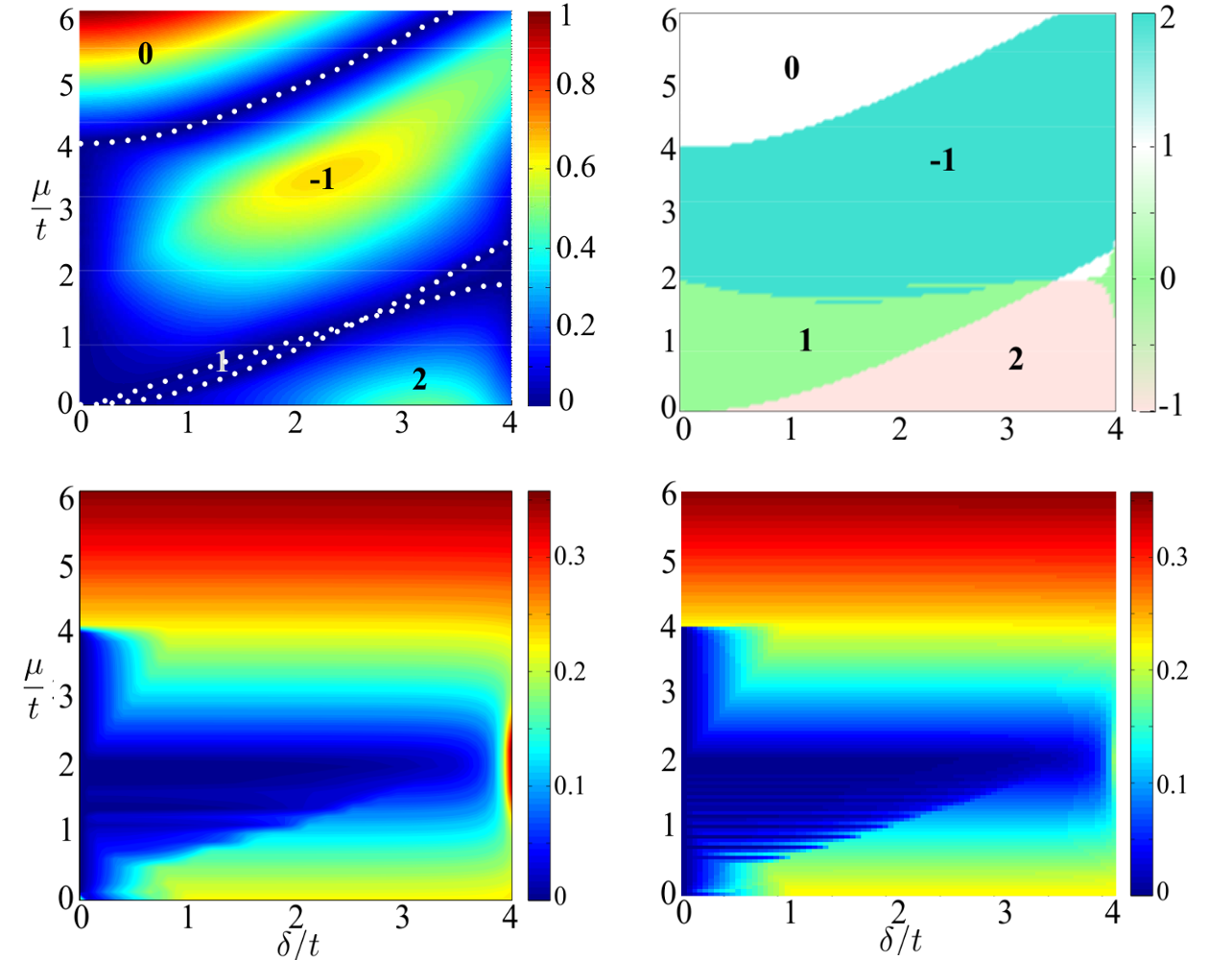}
	\caption{\emph{Top Left}: The phase diagram as computed via the Berry phase. The colour encodes the magnitude of the spectral gap while the dashed lines indicate the phase boundaries. \emph{Top Right}: The phase diagram as a sum of the winding numbers of the $a$ and $b$ sublattices. \emph{Bottom Left}: The sublattice entanglement as characterised by $\text{min}_{\pp}|\sv_{i}(\pp)|$. When $\text{min}_{\pp}|\sv_{i}(\pp)|$ is small the entanglement is large.  \emph{Bottom Right}: The gap of the entanglement spectrum corresponding to either sublattice. Here, the gap is defined to be $\text{min}_{\pp} |1/2 - \lambda_+| = \text{min}_{\pp} |1/2 - \lambda_-|$. For superconducting systems, the particle-hole symmetry of the entanglement spectrum, $\epsilon_+ = -\epsilon_-$, implies that $1/2 - \lambda_+ = -\left( 1/2 - \lambda_-\right)$ and gap closure corresponds to $\epsilon_+ = \epsilon_- = 0$. There is good correspondence between $\sv_{i}(\pp) \to 0$ and the gap closing.}
	\label{f1}
\end{figure}

We partition the system based on the sublattices, which implies that the relevant operators to evaluate the corresponding subsystem winding numbers are given by \eqref{eq:supobs}. Calculating the observable vectors $\s_{a}(\pp)$ and $\s_{b}(\pp)$ through evaluation of \eqref{equation:expb} and \eqref{equation:expb1}, Figure \ref{f1} shows that the phase diagram is in general faithfully reproduced. Discrepancies between the Chern number and the sum of the sublattice winding numbers occur only in regions where the sublattices become (close to) maximally entangled. Thus as we analytically argued above, caution should be taken when trusting the results in these regimes. Still, we note that the discrepancies are only in the signs, which means that all distinct types of topological phases are distinguished. 

%Comparing $\text{min}_{\pp}|\sv_{i}(\pp)|$ to the entanglement spectrum there is a clear correspondence between them. This confirms $|S|$ as a good measure of the entanglement between the two subsystems. 

\section{Experimental Applications}

We now briefly discuss how one might actually measure the subsystem winding numbers in a cold atom setup. We present the measurement scheme described in \cite{PALG13} for the model presented in example II. Given some lattice model with two sublattice degrees of freedom, one can model it in a system of cold atoms trapped in an optical lattice. Tunnelling amplitudes and chemical potentials can be implemented through Raman-assisted tunnelling \cite{JZ03,GD10,MBGRML12}. Through the use of two atomic states and $s$-wave Feshbach resonances \cite{JKAAP11,BP04} we can implement nearest neighbour pairing. Such a model supports topologically non-trivial phases.

In order to access the subsystem winding numbers we must measure the components of the vectors $\s_{i}(\pp)$, for $i=a,b$. The two sublattices are differentiated by their chemical potentials, allowing us to release them from the trap one at a time. The observables \eqref{eq:supobs} can then be measured independently. This technique uses the fact that time-of-flight images give direct access to the momentum space densities $\braket{a_{\pp}^{\da}a_{\pp}}$ and $\braket{b_{\pp}^{\da}b_{\pp}}$ which immediately returns $\hat{s}_{i}^{z}(\pp)$. We can retrieve the $\hat{s}_{i}^{x}(\pp)$ and $\hat{s}_{i}^{y}(\pp)$ components by switching off the pairing and tunnelling before releasing the atoms from the trap, which is equivalent to rotating the observables $\Sigma_{i}^{x,y}$ to $\Sigma_{i}^{z}$. One then performs time-of-flight measurements as before.

The reason this method is successful is because the winding number is linked to the physical observables given in \eqref{eq:supobs}. In principle, any two-species system in which these observables are accessible should be experimentally assessable for topological order. For example, angle-resolved photoemission spectroscopy has proven successful in measuring both spin textures \cite{Hsieh13022009} and orbital textures \cite{YW13} in condensed matter systems. 

Real physical systems are prone to disorder. In such instances, the reciprocal space represenatation of the winding number \eqref{equation:winding} no longer holds as translational invariance is broken. However, the winding number is still well defined in terms of real space observables \cite{K06}. Indeed the two represenations are equivalent under translational invariance. One can consider a translationally invariant system into which disorder in introduced. As long as the system remains in the same topological phase the winding numbers, in principle, remain accessible.

%The role of the winding number of a coupled pseudospin degree of freedom may also be relevant for ARPES-measurements in some families of topological insulators (Zhang et al. arXiv:1211.0762). Other example of a successful spin texture measurement method is spin-resolved photoemission (Eremeev et al. Nature Communications 3, Article number: 635). 

\section{CONCLUSIONS}
\label{sec:conclusions}

We have presented a method to detect Chern numbers of topological four-component insulators and superconductors. This method is based on an analytic decomposition of the Chern number as the sum of subsystem winding numbers, which in turn can be expressed in terms of the expectation values of observable quantities. When the subsystems were associated with spin components in the quantum spin Hall effect and sublattice degrees of freedom in a staggered superconductor, we showed that the phase diagrams of both models were accurately reproduced. Our method can be viewed as a generalization of the detection of Chern numbers from time-of-flight images \cite{WST13,PALG13,AFMPG11}. As such it is particularly well suited for cold atom experiments, where multiple internal atomic states are often used to synthesise the pairing terms, the spin-orbit couplings and gauge fields required for topological phases to emerge \cite{LP13,BGKLM10,GSNBMLS10,BC11,MBGRML12,BMRGLM10,KML10}.  Being able to separately measure the time-of-flight images of the components is sufficient to construct the subsystem winding numbers and hence the full Chern number.

The accuracy of our method is limited only by the entanglement between the components with respect to which the observables are defined. We found that the Brillouin zone discretisation errors become more significant as the components get more entangled and thus the component winding numbers, while still remaining integers, become more unreliable. Fortunately, we showed that the observables that we employ to construct the winding numbers can also probe the entanglement and thus assess the reliability of the results. In our examples, all discrepancies could be explained in terms of maximal entanglement occurring in the corresponding regions of the parameter space. While our decomposition of the Chern number in terms of the component winding numbers makes no  \textit{a priori} assumptions on the physical nature of the components, i.e. a decomposition exists for any partition, for reliable experimental application of our method, one should thus employ observables associated with components that are not maximally entangled. The observed low entanglement between spin components in the QSH insulator can also be viewed as a complementary argument for the robustness of the spin Chern number in the presence of the spin mixing Rashba term \cite{SWSH06}.

We also analytically showed that the entanglement between the components was directly related to the gap in the translation symmetry preserving component entanglement spectrum \cite{LN13}, with gap closures corresponding to maximally entangled modes. As the component entanglement is a physical observable in itself, our work provides a rare example of entanglement properties that can be probed through physical measurements. It is an interesting future direction to study what other entanglement (spectrum) properties, in particular those related to other types of system partitions \cite{LH08,SRB11,HF13}, could be accessed through measurements. 

Another open question is the generalisation of our analytic argument for the Chern number decomposition to $n$ component systems. While Schmidt decompositions are hard to generalise for systems with more than two components \cite{AACJLT00}, a possible avenue might be to use convoluted bi-partite Schmidt decompositions.

\section{ACKNOWLEDGEMENTS}

\noi This work was supported by EPSRC, Spanish MINECO Project FIS2012-33022, Beca FPU No. AP 2009-1761, and CAM research consortium QUITEMAD S2009-ESP-1594. VL is supported by the Dutch Science Foundation NWO/FOM.

\bibliographystyle{prs}
\bibliography{ref}

\newpage

\appendix
\section{Berry phase and winding number for the 2-dimensional case}
\label{App:Chernreps}

Here we look into the different representations of the Chern number. In particular, we present the relation between the Berry phase expression of the Chern number and its winding number expression for the case where the kernel Hamiltonian is 2-dimensional.

\subsection{The Chern Number}

\noi Consider the Hamiltonian of a 2-dimensional many body system of non-interacting fermions given by

\begin{equation}\label{ham11}
H=\int_{\pp}d^{2}p\,\,\PSI^{\da}_{\pp}h(\pp)\PSI_{\pp},
\end{equation}

\begin{figure}[h!]
	\center	
	\includegraphics[scale=0.25]{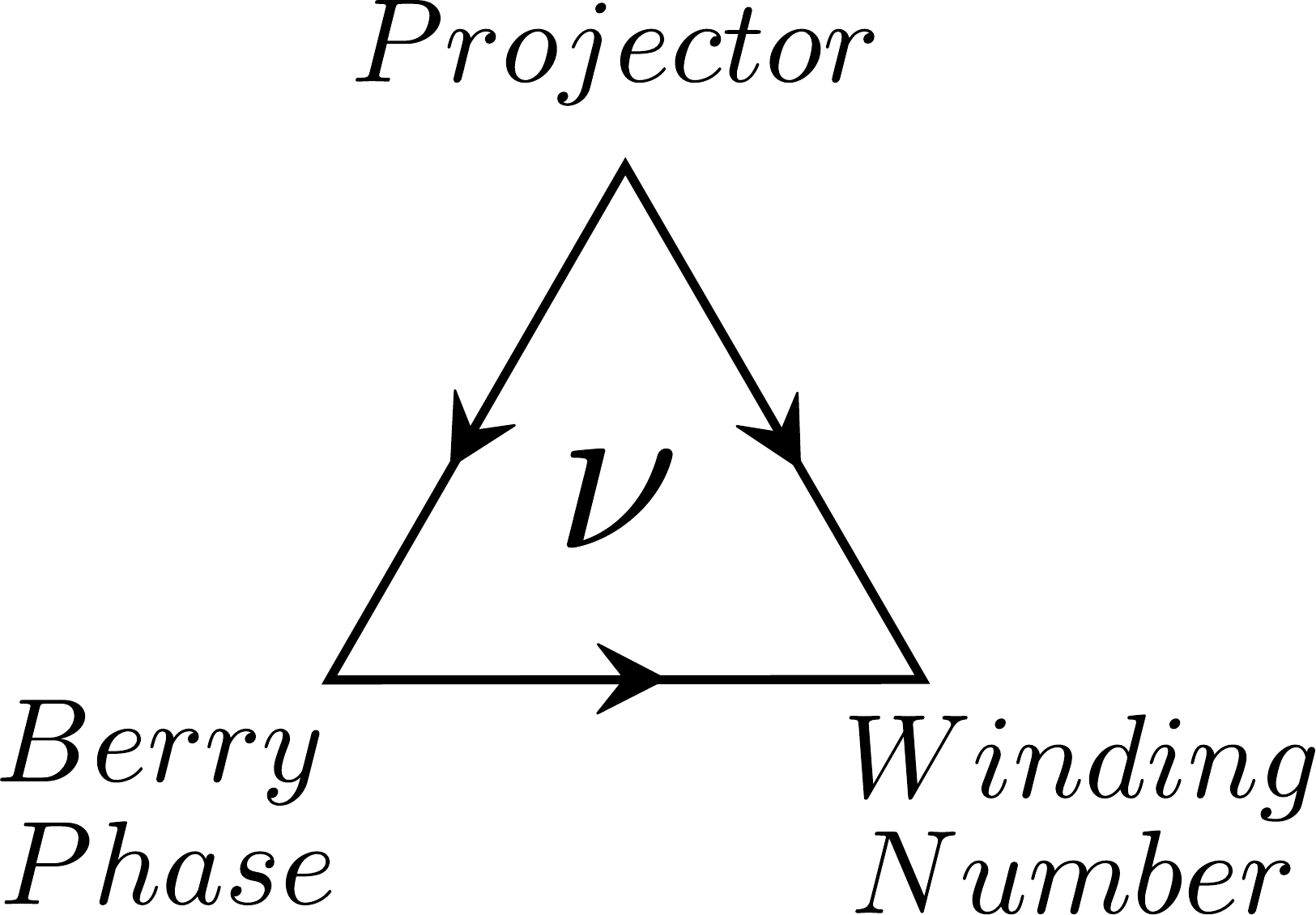}
	\caption{A schematic picture of the different representations of the Chern number, $\nu$. The arrows indicate the direction of the proofs given in this section.}
	\label{fig:tri}
\end{figure}

\noi where $\pp=\bpm p_{x}, & p_{y}\epm$ is an element of the Brillouin zone, $h(\pp)$ is the Bloch Hamiltonian and $\PSI_{\pp}$ is the two dimensional spinor containing the relevant field mode operators. We denote the ground state of the Bloch Hamiltonian by $\ket{\psi(\pp)}$ . Let us define a set of projectors $P_{\pp}=\ket{\psi(\pp)}\bra{\psi(\pp)}$. The Chern number $\nu$, of the first Chern class, is one of many topological invariants that define an order that cannot be destroyed by local unitary transformations. The Chern number has three representations of interest, illustrated in Fig. \ref{fig:tri}.

\noi The most fundamental representation is in terms of projectors $P_{\pp}$. It is written as
\begin{equation}\label{eqn:chernprojapp}
\nu=-\frac{i}{2\pi}\int_{BZ}d^{2}p \,\,\text{tr}\left(P_{\pp}\left[\partial_{p_{x}}P_{\pp},\partial_{p_{y}}P_{\pp}\right]\right).
\end{equation}

\noi From this expression we can retrieve both the Berry phase and winding number representations. 

\subsection{The Berry Phase}

\noi We can write the Chern number as the Berry phase that the ground state accrues around the boundary of the Brillouin zone. To see this we take the projector representation \eqref{eqn:chernprojapp} and substitute in the definition $P_{\pp}=\ket{\psi(\pp)}\bra{\psi(\pp)}$
\begin{equation}\label{eqn:berrychern}
\nu=-\frac{i}{2\pi}\int_{BZ}d^{2}p\bra{\psi(\pp)}\left[\ket{\partial_{p_{x}}\psi(\pp)}\bra{\psi(\pp)}+\ket{\psi(\pp)}\bra{\partial_{p_{x}}\psi(\pp)},\ket{\partial_{p_{y}}\psi(\pp)}\bra{\psi(\pp)}+\ket{\psi(\pp)}\bra{\partial_{p_{y}}\psi(\pp)}\right]\ket{\psi(\pp)}.
\end{equation}

\noi The ground state is normalised so we have $\partial_{\mu}\braket{\psi(\pp)|\psi(\pp)}=0$, where $\mu=p_{x},p_{y}$, such that
\begin{equation}\label{eqn:anti}
\braket{\partial_{\mu}\psi(\pp)|\psi(\pp)}=-\braket{\psi(\pp)|\partial_{\mu}\psi(\pp)}.
\end{equation}

\noi Expanding \eqref{eqn:berrychern} and using relation \eqref{eqn:anti} we obtain
\begin{equation}
\nu=-\frac{i}{2\pi}\int_{BZ}d^{2}p\left(\braket{\partial_{p_{x}}\psi(\pp)|\partial_{p_{y}}\psi(\pp)}-\braket{\partial_{p_{y}}\psi(\pp)|\partial_{p_{x}}\psi(\pp)}\right).
\end{equation}

\noi We now recognise that the integrand is the Berry curvature, $F_{p_{x}p_{y}}$,
\begin{equation}
F_{p_{x}p_{y}}=\partial_{p_{x}}A_{p_{y}}-\partial_{p_{y}}A_{p_{x}}=\braket{\partial_{p_{x}}\psi(\pp)|\partial_{p_{y}}\psi(\pp)}-\braket{\partial_{p_{y}}\psi(\pp)|\partial_{p_{x}}\psi(\pp)},
\end{equation}

\noi where $\boldsymbol{A}=\braket{\psi(\pp)|\PA|\psi(\pp)}$ is the Berry connection. Using Stokes' theorem, the Chern number can be written 
\begin{equation}
\nu=-\frac{i}{2\pi}\int_{BZ}d^{2}p\,\,F_{p_{x}p_{y}}=-\frac{i}{2\pi}\oint_{\partial BZ}d\pp\cdot\boldsymbol{A},
\end{equation}

\noi where $\partial BZ$ is the boundary of the Brillouin zone. Hence, the Chern number can be written in terms of the Berry phase accumulated by the ground state around the boundary of the Brillouin zone.

\subsection{The Winding Number}

\noi When the Bloch Hamiltonian $h(\pp)$ is 2-dimensional the ground state $\ket{\psi(\pp)}$ lies on the Bloch sphere. Therefore we can write down an explicit form of the projector in terms of a normalised 3-vector $\s(\pp)$
\begin{equation}\label{windproj}
P_{\pp}=\ket{\psi(\pp)}\bra{\psi(\pp)}=\frac{1}{2}\left(1\!\!1-\s(\pp)\cdot\SIG\right),
\end{equation} 

\noi where $1\!\!1$ is the identity matrix and $\SIG=\bpm\sigma^{x},&\sigma^{y},&\sigma^{z}\epm$ is the vector of Pauli matrices. Substituting the preceeding definition into the definition of the Chern number in terms of projectors \eqref{eqn:chernprojapp} and  employing the identities
\begin{equation}\label{vecid}
(\boldsymbol{a}\cdot\SIG)(\boldsymbol{b}\cdot\SIG)=1\!\!1\boldsymbol{a}\cdot\boldsymbol{b}+i\SIG\cdot\boldsymbol{a}\times\boldsymbol{b},
\end{equation}

\noi for two 3-vectors $\boldsymbol{a}$ and $\boldsymbol{b}$, and
\begin{equation} 
\text{tr}\left(\sigma^{\alpha}\sigma^{\beta}\right)=2\delta^{\alpha\beta},\qquad \alpha,\beta=x,y,z,
\end{equation}

\noi we obtain
\begin{align}\label{eqn:windchern}
\nu&=-\frac{i}{2\pi}\int_{BZ}d^{2}p\,\,\text{tr}\left(P_{\pp}\left[\partial_{p_{x}}P_{\pp},\partial_{p_{y}}P_{\pp}\right]\right)\nn\\
&=-\frac{i}{2\pi}\int_{BZ}d^{2}p\,\,\text{tr}\left(\ket{\psi(\pp)}\bra{\psi(\pp)}\frac{1}{4}\left[\partial_{p_{x}}\s(\pp)\cdot\SIG,\partial_{p_{y}}\s(\pp)\cdot\SIG\right]\right)\nn\\
&=\frac{1}{4\pi}\int_{BZ}d^{2}p \,\, \s(\pp)\cdot\left(\px\s(\pp)\times\py\s(\pp)\right).
\end{align}

\begin{figure}[h!]
	\center	
	\includegraphics[scale=0.5]{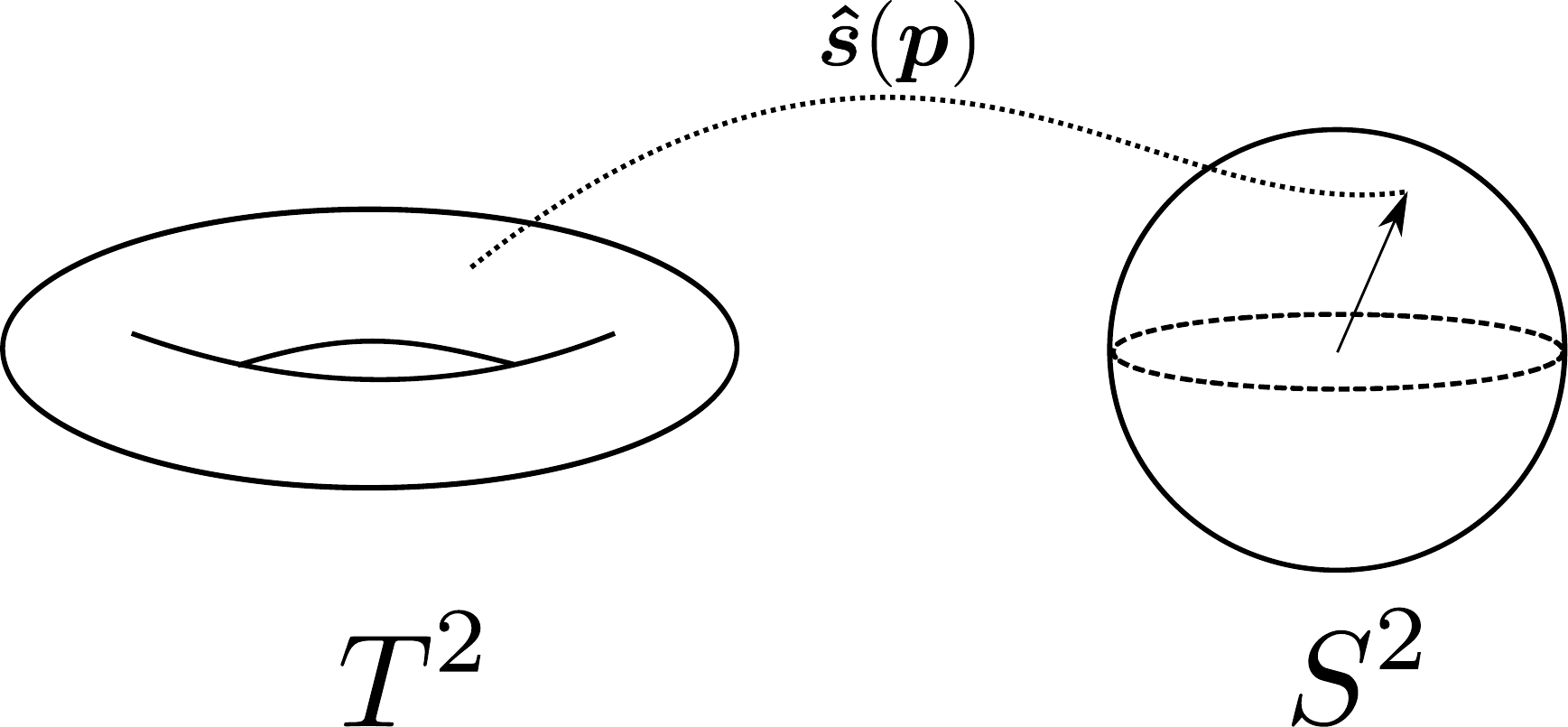}
	\caption{The winding number describes the number of times the map $\s(\pp):T^{2}\rightarrow S^{2}$ wraps around the unit 2-sphere.}
	\label{fig:wind}
\end{figure}

\noi The vector $\s(\pp):T^{2}\rightarrow S^{2}$ is a map between the toroidal Brillouin zone and the unit 2-sphere as shown in Fig. \ref{fig:wind}. The expression for the Chern number derived above \eqref{eqn:windchern} describes the winding of this map around the unit sphere when the Brillouin zone is spanned.  

\subsection{Berry Phase to Winding Number}

\noi To complete the picture outlined in Fig. \ref{fig:tri} we now derive the winding number directly from the Berry phase. We start by writing the Chern number as the integral of the Berry curvature
\begin{equation}\label{btw}
\nu=-\frac{i}{2\pi}\int_{BZ}d^{2}p\left(\braket{\partial_{p_{x}}\psi(\pp)|\partial_{p_{y}}\psi(\pp)}-\braket{\partial_{p_{y}}\psi(\pp)|\partial_{p_{x}}\psi(\pp)}\right)
\end{equation}

\noi Assuming the Bloch Hamiltonian is 2-dimensional we can write the projector as \eqref{windproj}. We substitute the projector into \eqref{btw}
\begin{equation}
\nu=-\frac{i}{2\pi}\int_{BZ}d^{2}p\left(\partial_{p_{x}}\left(\bra{\psi(\pp)}P_{\pp}\right)\partial_{p_{y}}\left(P_{\pp}\ket{\psi(\pp)}\right)-\partial_{p_{y}}\left(\bra{\psi(\pp)}P_{\pp}\right)\partial_{p_{x}}\left(P_{\pp}\ket{\psi(\pp)}\right)\right).
\end{equation}

\noi Using the fact that $P^{2}=P$, $\left\{\partial_{\mu}\s(\pp)\cdot\SIG,\s(\pp)\cdot\SIG\right\}=0$ and relation \eqref{vecid} we find that
\begin{align}
\nu &= \frac{1}{4\pi}\int_{BZ}d^{2}p\,\,\braket{\psi(\pp)|\SIG\cdot\left(\partial_{p_{x}}\s(\pp)\times\partial_{p_{y}}\s(\pp)\right)|\psi(\pp)}\nn\\
&=\frac{1}{4\pi}\int_{BZ}d^{2}p \,\, \s(\pp)\cdot\left(\px\s(\pp)\times\py\s(\pp)\right).
\end{align}

\end{document}